\documentstyle[prb,aps,psfig]{revtex}

\newcommand{\eq}[2]{\begin{equation} #1 \label{eq:#2} \end{equation}}

\newcommand{\req}[1]{(\ref{eq:#1})}

\begin{document}

\draft

\title{LANDAU MODEL FOR UNIAXIAL SYSTEMS WITH COMPLEX ORDER PARAMETER}

\author{M. Latkovi\'{c} and A. Bjeli\v{s}}

\address{
 Department of Theoretical Physics, Faculty of Science \\
 University of Zagreb,  P.O.B. 162, 10001 Zagreb, Croatia }

\maketitle

\begin{abstract}
We study the Landau model for uniaxial incommensurate-commensurate
systems of the I class by keeping Umklapp terms of third and fourth 
order in the expansion of the free energy. It applies to systems in 
which the soft mode minimum lies between the corresponding commensurate 
wave numbers. The minimization of the Landau functional leads to the 
sine-Gordon equation with two nonlinear terms, equivalent to the 
equation of motion for the well-known classical mechanical problem of 
two mixing resonances. We calculate the average free energies for
periodic, quasiperiodic and chaotic solutions of this equation,
and show that in the regime of finite strengths of Umklapp terms
only periodic solutions are absolute minima of the free energy,
so that the phase diagram contains only commensurate configurations.
The phase transitions between neighboring configurations are of the
first order, and the wave number of ordering goes through harmless
staircase with a finite number of steps. These results are the basis 
for the interpretation of phase diagrams for some materials from 
the I class of incommensurate-commensurate systems, in particular 
of those for A$_2$BX$_4$ and BCCD compounds. Also, we argue that 
chaotic barriers which separate metastable periodic solutions 
represent an intrinsic mechanism for observed memory effects and 
thermal hystereses. 

\end{abstract}


\section{Introduction}

Usual treatments of uniaxial incommensurate-commensurate (IC-C)
phase transitions are based either on microscopic models with
competing interactions or on phenomenological Landau theories.
The relevant reviews can be found in Refs.~\cite{janssen,neubert}.
The well-known example of the former is the Frenkel-Kontorova (FK)
model~\cite{aubr,bak}, in which the wave number of ordering goes
through the devil's staircase sequence of second order phase 
transitions~\cite{aubr}. In the regime of weak interactions the
FK model can be continuated, and so reduced to the exactly solvable
(i.~e. integrable) sine-Gordon model~\cite{bak}. The solutions
which then participate in the phase diagram are phase soliton
lattices, i.~e. commensurate regions separated by so called
discommensurations~\cite{mcmi}. The phase transition to the
commensurate state is of the second (continuous) order, and the
devil's staircase variation of the wave number is replaced by its
simple continuous dependence on the control parameter.

The phenomenological Landau theory, another usual approach to the
IC-C transitions, starts from the expansion of the thermodynamic
potential in terms of the order parameter, relied on the symmetry
requirement by which the order parameter is defined through one of
the irreducible representations of the symmetry group of the normal
phase. E.~g., for structural phase transitions the order parameter
is defined as a set of normal coordinates of the soft
mode~\cite{cowl,tole}. Generally the minimum frequency
of this soft mode may be located at an arbitrary point (i.~e.
star of wave vectors) in the first Brillouin zone. The simplest
irreducible representation for an uniaxial ordering is then
two-dimensional. The corresponding basic ("minimal") form of the
Landau expansion comprises, besides the leading normal terms, one,
presumably the strongest, Umklapp term allowed by symmetry. This
term favors a commensurate ordering and is responsible for the
lock-in transition from the incommensurate ordering favored by
the elastic term. Minimization of the Landau functional again
leads, after neglecting the space variations of the order
parameter amplitude~\cite{mcmi}, to the sine-Gordon
equation~\cite{emery,bula}, i.~e. to the phase diagram equal to
that of the FK model after the space continuation.

The above approaches predict either a dense sequence of second order
phase transitions (devil's staircase in the FK model) or an isolated
transition of the same type (Landau theory). Both possibilities are
indeed close to the observations of IC-C transitions in some
materials~\cite{cumm,blinc}. A majority of materials however exhibits
a more complex behavior comprising one or more first order phase
transitions, memory effects, wide ("global") hystereses, finite density
of solitons at the very IC-C transition, etc (for a review see e.~g.
Ref.~\cite{blinc}). It is usually difficult
to decide solely from the experimental observations, even for the most
carefully prepared samples, whether such effects are of purely intrinsic
or of some extrinsic origin. From the theoretical side, they cannot be
explained within either of above approaches without extending the models.
So far this problem was mainly considered by taking primarily into
account some extrinsic agents, like external fields (e.~g. electric field
in ferroelectric materials), pinning centers, fixed or mobile defects,
additional external periodic potentials with a periodicities different
from that already present in the model), etc. 

Another, more intricate possibility is that of intrinsic sources and
mechanisms as the potential explanations for the afore mentioned
phenomena~\cite{bari}. In this respect the central question is the
following; what are the simplest intrinsic extensions of the above
basic approaches which lead to phase diagrams with a finite sequence
of first order transitions (i.~e. harmless staircase~\cite{vill}), 
and thus offer an inherent explanation for global hystereses and
corresponding phenomena. 

The attempts in this direction were more successful in the realm
of discrete models. The examples are models which include couplings
between next nearest neighbors, like so-called DIFFOUR
model~\cite{janssen1}, axial next-nearest neighbor Ising (ANNNI)
model~\cite{fisher,selke1}, as well as various
extensions~\cite{yama,selke2,selke}, and models with two spin-like 
variables per site like those of Chen and Walker~\cite{chen} and 
Janssen~\cite{janssen2}. Both types of extensions were
aimed mostly towards the interpretation of the phase
diagrams observed in the family of A$_2$BX$_4$ compounds.  

On the other hand, the attempts within Landau models were based on 
the formal inclusion of more and more Umklapp terms (i.~e. stars of 
wave vectors) into the basic models for the classes
I~\cite{mashi,marion,parl3} and II~\cite{ribeiro,sannikov1,sannikov2}
of IC-C systems. From one side, the relevance of the Umklapp terms of
high orders in the Landau expansion can be hardly justified on the
physical grounds. Also, the ensuing analyses took into account only
sinusoidal modulations, which, as the present study shows, is a too
crude approximation for the determination of phase diagrams with
harmless staircase, as well as for the interpretation of accompanying
hysteretic effects.  

In contrast to such approaches, we propose in the present work a simple,
physically well justified, extension of the basic Landau model for the 
class I, which is still framed within a "minimal" free energy expansion
for a single star of wave vectors. 
The phase diagram which emerges from our model is characterized by a
harmless staircase and first order transitions between highly nonsinusoidal
configurations with different periods. Furthermore, a closer examination 
of configurations that participate in the phase diagram, and also of 
those that are not thermodynamically favored, enable 
a plausible explanation of the memory and hysteresis effects as the
intrinsic (or at least semi-intrinsic) properties of IC-C systems.

Our considerations are based on a sine-Gordon model with two Umklapp
terms~\cite{bjel}. This type of model is physically well-grounded
whenever the Landau expansion contains terms which favor two different
commensurabilities, that are of comparable strengths.
The most interesting case is realized with Umklapp terms of third and
fourth order, the lowest possible ones within the models with the
Lifshitz invariant, appropriate for the so-called systems of the I
class~\cite{cowl} (the systems of the II class have lock-in
transitions at the commensurabilities of order one and two, and are
covered by essentially different type of Landau models~\cite{mich,dana}).

The mean-field (saddle point) approximation for our Landau functional
leads to the Euler-Lagrange (EL) equation which has the form of the
double sine-Gordon equation. This is one of the most intensively
studied nonintegrable problems in the contemporary classical
mechanics~\cite{chir,esca,zasl}. The corresponding phase portrait
contains periodic, quasiperiodic and chaotic trajectories, the latter
appearing only when {\em both} nonlinear terms in the Landau functional 
are finite. As the strength of nonlinear terms increases, the chaotic
trajectories occupy larger and larger portion of the phase space,
destroying gradually quasiperiodic Kolmogorov-Arnold-Moser (KAM)
layers, and eventually allowing only for some isolated periodic
trajectories. The latter are orbitally unstable and therefore are not
realized within the scope of classical mechanics. However, we show
that just this tiny subset of the phase space comprises local minima
of the free energy functional, i.~e. the solutions (configurations)
which participate in the thermodynamic phase diagram. The question
which then arises is analogous to that met in the analyses of the
discrete models~\cite{aubr,frad}, i.~e. are there thermodynamically
stable configurations among other, quasiperiodic and chaotic,
trajectories.

In order to analyze this additional, {\em thermodynamic},
aspect of the phase portrait, we calculate the average free energy for
periodic, quasiperiodic and a representative set chaotic solutions
of EL equation, with the aim to find, for given values of control
parameters, those solutions that have the lowest value of the average
free energy. We show that the chaotic configurations are never
thermodynamically stable, in agreement with results obtained for some
discrete models~\cite{aubr,frad}. The quasiperiodic configurations
might be present in the phase diagram only
when the Umklapp terms are weak enough, i.~e. at temperatures slightly
below the phase transition from the disordered to the incommensurate
state. In the regime of strong Umklapp terms (to be specified later)
the phase diagram is completely covered by periodic configurations,
and the wave number of ordering passes through a finite number of
values separated by the first order transitions, i.~e. the
corresponding staircase is harmless.

The article is organized as follows. In Sec. II we introduce
the Landau model of uniaxial ordering with two Umklapp terms and
discuss its classical mechanical counterpart. The solutions
of the Euler-Lagrange equation are considered in Sec. III,
and the corresponding thermodynamic phase diagrams are presented
in Sec. IV. Finally, in Sec. V we discuss possible implications to 
the phenomena observed in real materials, and compare our results 
with those obtained in the previous analyses of the similar models
and other theories of uniaxial IC-C ordering.


\section{Model}

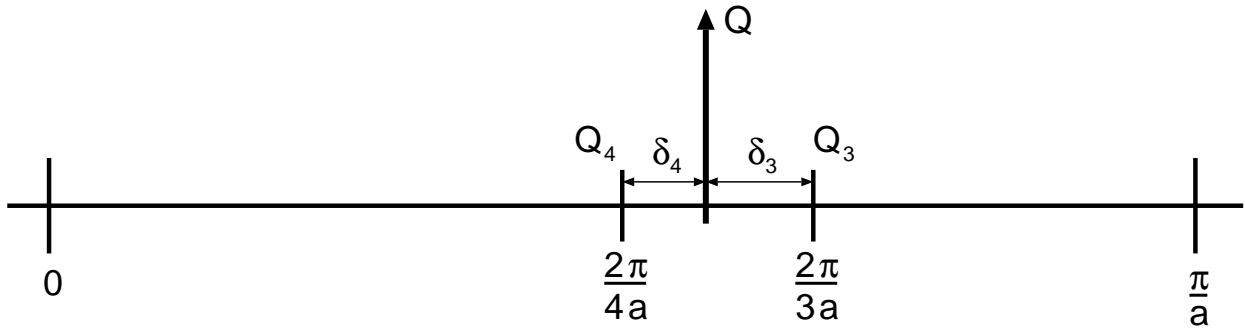
\begin{figure}
\caption{Brillouin zone with the soft-mode minimum at $Q$, 
and the commensurate wave numbers  of third ($Q_3$)
and fourth ($Q_4$) order.}
\label{vector}
\end{figure}

We start from the assumption that the quadratic contribution to
the Landau expansion has minima at wave numbers $(+Q,-Q)$, where
$Q_4<Q<Q_3$, with $Q_4=2\pi/4$ and $Q_3=2\pi/3$. Here the unit length
is taken equal to the lattice constant. The distances of $Q$ 
from $Q_3$ and $Q_4$ are denoted by $\delta_3$ and $\delta_4$
respectively, with $\delta_3+\delta_4=\pi/6$ (Fig.~\ref{vector}).
From now on we shall use $\delta_4$ as an independent control
parameter. Let us furthermore specify that the order parameter is
complex, $\rho e^{i\phi}$. Limiting the further analysis to the
temperature range well below the critical temperature for the
transition from the disordered to the incommensurate phase, we
also make the usual approximation of space independent amplitude 
$\rho$~\cite{mcmi}, and keep only the phase-dependent part of the
free energy density. The latter reads
\eq{ f(\phi,x)=\frac{1}{2}(\frac{d\phi}{dx})^2+
	B\cos{[3\phi+3(\frac{\pi}{6}-\delta_4)x]}+
	C\cos{(4\phi-4\delta_4 x)}.}{free} 
Here we scale the free energy functional
\eq{F = \int dx f(\phi,x) ,}{Free} 
by $\xi_{0}^2 \rho^2$, where $\xi_0$ is the correlation length in the 
$x$-direction. The first, gradient term in Eq.~\req{free} is the
elastic contribution which favors the incommensurate sinusoidal
ordering with the wave number $Q$. The second and third terms are
the Umklapp contributions of the third and fourth order respectively. 
Due to the closeness of the wave number $Q$ to both respective 
commensurate wave numbers, they are presumably the leading 
Umklapp contributions, provided both are allowed by symmetry.
Their effective strengths are denoted by coefficients $B$ and $C$
which are proportional to the first and the second power of the amplitude
$\rho$ respectively. They are another two control parameters (beside
$\delta_4$) of the model~\req{free}. The temperature variation of
$\rho$ is expected to be the main source of the temperature dependence
of $B$ and $C$.

The model~\req{free} covers a
variety of possibilities which may take place in particular 
physical examples. Besides the competition of each Umklapp term
and the elastic term present already in basic sine-Gordon models,
the essential new property of the present model is an
additional competition between two Umklapp terms. The relative
importance of these two terms is relied on both, the ratio of
the strengths $B$ and $C$ and the position of the wave number $Q$,
i.~e. the ratio of $\delta_3$ and $\delta_4$, so that various
regimes are possible. Regarding the expansion~\req{free} it is
reasonable to expect that the relative strength of two terms varies
from the dominance of the third order term ($B\gg C$) at temperatures
not far below the transition from the disordered phase to the
comparable values of $B$ and $C$ at lower temperatures. However,
even when $B\gg C$, the relative weakness of the fourth order Umklapp 
term can be compensated by a small value of $\delta_4$ with respect
to $\delta_3$, i.~e. by its much slower space dependence. In this
case it is necessary to keep both Umklapp terms in the
expansion~\req{free}. Although similar arguments may be invoked
in favor of retaining some other pair of commensurate wave 
numbers, or even more than two of them, the example~\req{free} seems 
to be the most interesting one, due to the lowest possible powers of 
$\rho$ present in the coefficients $B$ and $C$. 

The configurations which take part in the thermodynamic phase
diagram of the model~\req{free} are the solutions of the
Euler-Lagrange (EL) equation,
\eq{ \phi''+3B\sin{[3\phi+3(\frac{\pi}{6}-\delta_4) x]}+
	4C\sin{(4\phi-4\delta_4 x)}=0 ,}{el}
which for given values of the control parameters have the lowest
value of the free energy averaged over the macroscopic length of
the system, $L$, 
\eq{ \langle F \rangle =\frac{1}{L}\int dx f[\phi(x), x] .}{mfe}

Before developing the appropriate method for the determination of 
such configurations, let us put few remarks on the equation~\req{el}.
From the classical mechanical side it represents the nonintegrable
double resonance (i.~e. double sine-Gordon) model~\cite{chir,esca},
with the corresponding Hamiltonian
\eq{ H(p_{\phi},\phi,x)=\frac{p_{\phi}^2}{2}-
	B\cos{[3\phi+3(\frac{\pi}{6}-\delta_4) x]}-
	C\cos{(4\phi-4\delta_4 x)} ,}{ham}
where $p_{\phi} \equiv \frac{\partial f}{\partial\phi'}=\phi'$.
Obviously for $B=0$ or $C=0$ Eqs.~\req{el} and~\req{ham} reduce to 
completely integrable sine-Gordon problems. For both $B$ and
$C$ finite, one encounters the coexistence of two overlapping
resonance domains. This can be easily seen with help of
Poincar\'{e} cross section. We introduce the auxiliary variable
\eq{ \psi=\phi+(\frac{\pi}{6}-\delta_4)x ,}{psi}
and plot the Poincar\'{e} cross section in the phase space
$(\psi,p_{\psi})$, $\psi\equiv\psi(x_0+3n), p_{\psi}\equiv\psi'(x_0+3n)$.
Here $x_0$ is the starting point of integration and $n$ is an integer.
The resonance domains are situated around elliptic fixed points
at $(\psi=0,p_{\psi}=\frac{2\pi}{3}m)$ and
$(\psi=\frac{\pi}{6},p_{\psi}=\frac{\pi}{4}(2m+1))$, where
$m$ is an integer. Their respective widths are $12\sqrt{B}/\pi$
if $C=0$ and $12\sqrt{C}/\pi$ if $B=0$. For small values
of $B$ and $C$ (Fig.~\ref{poinc}a) the trajectories between two
resonances conserve their topological form, while chaotic
trajectories exist only very close to the separatrices of both
resonances. As $B$ and/or $C$ increase (Fig.~\ref{poinc}b) the
separatrices burst out into stochastic layers which grow and
eventually merge between resonance domains. One gradually arrives
at the threshold of the stochasticity (Fig.~\ref{poinc}c),
given by the Chirikov criterion~\cite{chir}
\eq{ \frac{12}{\pi}(\sqrt{B}+\sqrt{C}) \approx 1 ,}{chirc}
at which the last KAM torus is destroyed, i.~e. there are no more
quasiperiodic solutions between two resonances.
Chaotic trajectories are now free to diffuse through all phase
space between two resonances.  Let us here mention two points relevant
for the further discussion. First, the widths of the chaotic layers
grow exponentially~\cite{mack} as parameters $B$ and $C$ increase.
Thus, the area between two resonances will be rapidly covered with
chaotic layers. Second, KAM tori represent the main
obstacles to diffusion of chaotic trajectories through phase
space (\cite{zasl} and references therein).


\section{Solutions of the Euler-Lagrange equation}

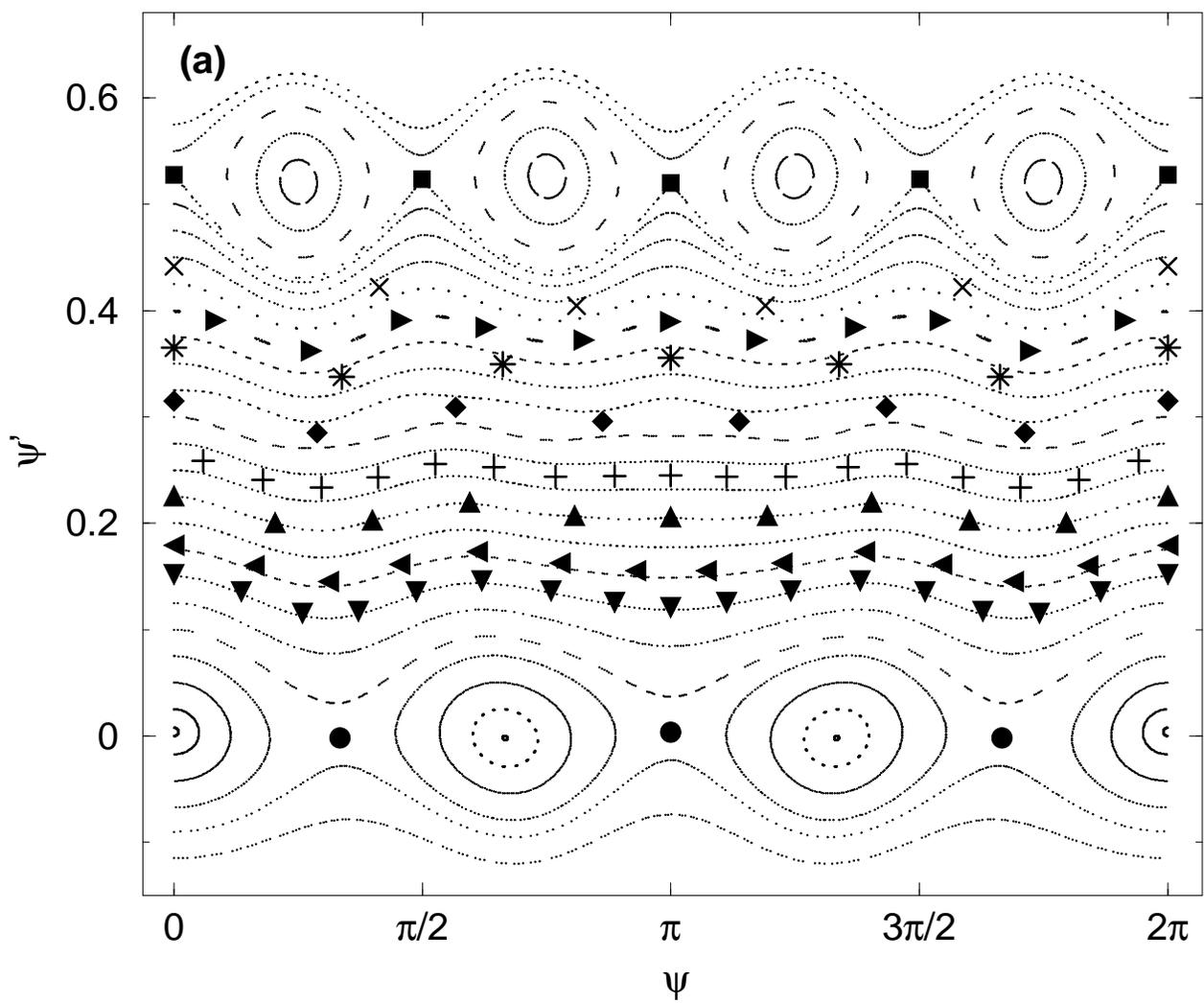
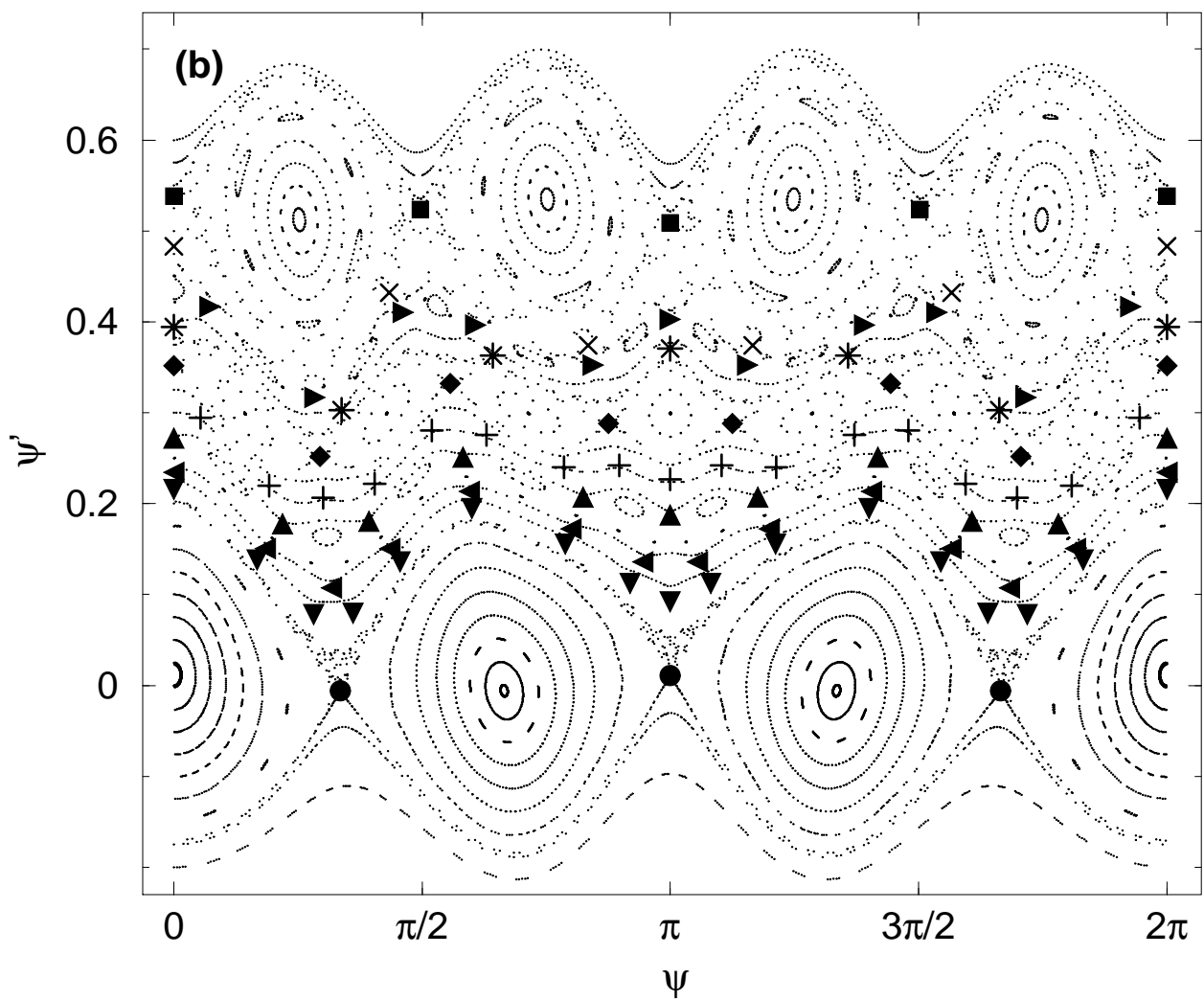
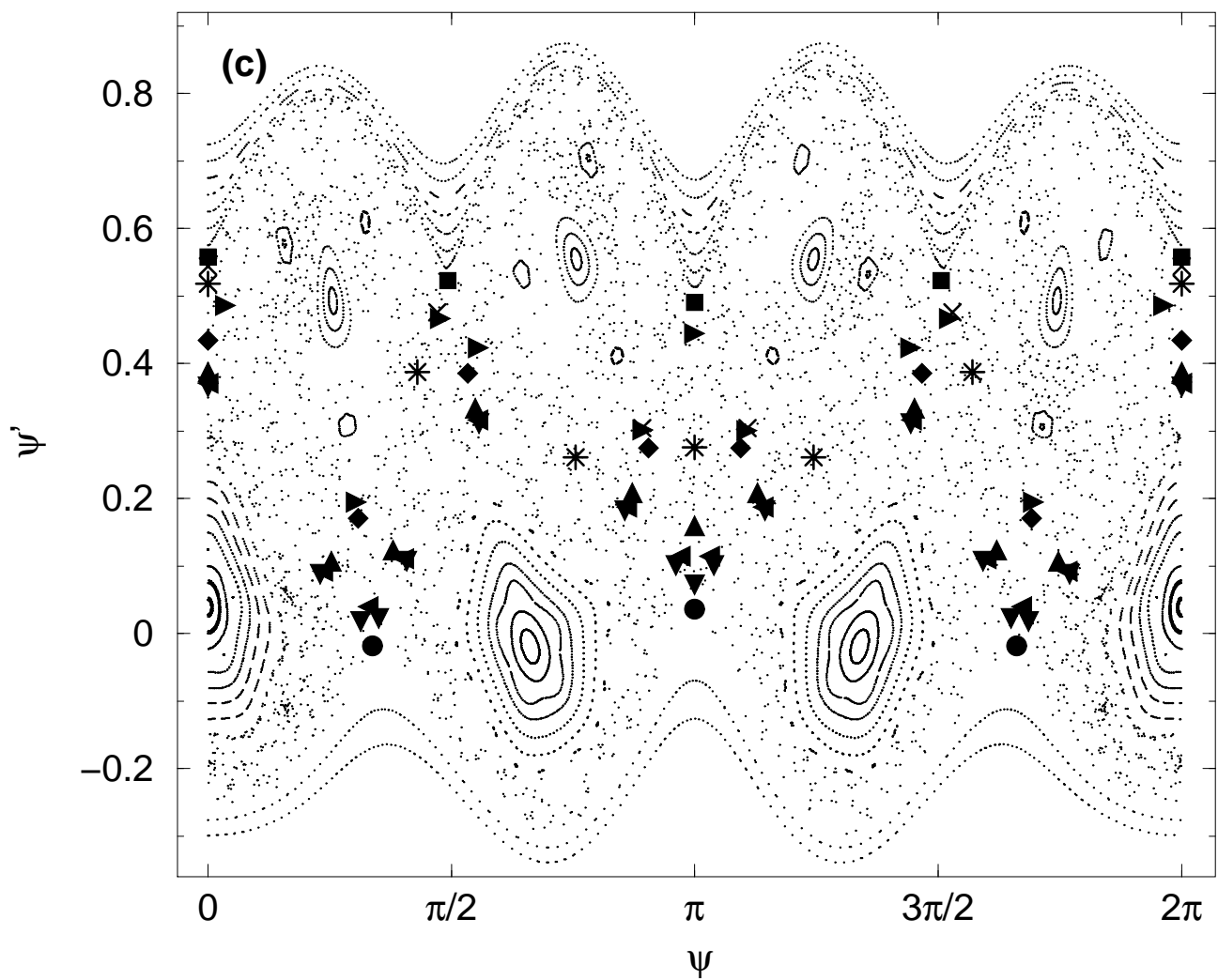
\begin{figure}
\caption{The Poincar\'{e} cross sections for the Euler-Lagrange
equation~\protect{\req{elpsi}} and the choice of parameters:
$x_0=1.5$, $\psi(x_0)=0$, $B=C=0.002$ (a), $B=0.008, C=0.006$ 
(b), $B=C=0.02$ (c). The period which defines the section is 
equal to 3. Symbols for the periodic solutions $(k, l)$ are:
solid $\bigcirc$         (0,1),
solid $\Box$             (1,0),
solid $\Diamond$         (1,1),
solid $\bigtriangleup$   (1,2),
solid $\lhd$             (1,3),
solid $\bigtriangledown$ (1,4),
solid $\rhd$             (2,1),
$+$                      (2,3),
$\times$                 (3,1),
$\ast$                   (3,2).}
\label{poinc}
\end{figure}

Beside the classical mechanical context, our problem has an 
additional aspect, namely we are looking for the thermodynamically 
stable solutions, i.~e. the trajectories in the phase space from 
Figs.~\ref{poinc}a,b,c which are local minima of the functional 
\req{Free}. Since we have to compare average free energies~\req{mfe}
of the trajectories present in the phase space, our first task is
to specify numerical methods appropriate for the calculation of
particular types of solutions. 

The orbitally unstable periodic solutions obviously cannot be
determined by a direct integration of EL equations~\req{el},
commonly used for calculation of  orbitally stable 
solutions. It is therefore necessary to calculate them by using
a suitable boundary value method for nonlinear equations. The
most natural choice is the finite difference method, which is 
however rather demanding regarding the computer memory and 
time. It is therefore important to reduce the search for periodic
solutions by establishing in an analytic way sufficient conditions on 
the possible values of periods. To this end we start from the relation
\eq{ \phi(x+P)=\phi(x)+\phi_P ,}{phiper}
which holds for any periodic solution. Here $P$ is its period
and $\phi_P$ is the phase increment per period (note that the
periodic solutions with finite $\phi_P$ belong to the rotational
part of the phase space). Inserting Eq.~\req{phiper} into the
EL equation~\req{el} taken at $x+P$ one gets
\eq{ \phi'' + 3B\sin{[3\phi+3(\frac{\pi}{6}-\delta_4)x + 
     3\phi_P + 3(\frac{\pi}{6}-\delta_4) P]} +
     4C\sin{(4\phi-4\delta_4 x + 4\phi_P - 4\delta_4 P )}=0,}{Pel}
where $\phi \equiv \phi(x)$. Sufficient conditions on the values
of parameters $P$ and $\phi_P$ follow from the requirement that
Eqs.~\req{Pel} and~\req{el} have the same form, i.~e. that
\eq{ 3 \phi_P + 3(\frac{\pi}{6}-\delta_4) P = 2\pi k,
 \,\,\,\, 4\phi_P - 4\delta_4 P = -2\pi l, }{kl}
where $k$ and $l$ are integers. Thus we get 
\eq{ P=4k+3l, \hspace*{1cm} \phi_P=\delta_4 P-l\frac{\pi}{2} .}{per}
Obviously, each periodic solution satisfying the requirement~\req{kl}
is uniquely defined by a pair of integers $(k,l)$ which do not have 
a common integer factor~\cite{bjel}. Note that the above procedure, 
in particular the step from Eq.~\req{Pel} to the conditions~\req{kl},
in principle does not forbid the existence of periodic solutions which
do not belong to the set defined by Eqs.~\req{per}. However, our
attempts to locate numerically such solutions, although based on 
two independent algorithms, present and the alternative one~\cite{dlb}, 
always led to a negative result. This is an indication that the
solutions with the periods~\req{per} are very probably the only
possible periodic solutions, i.~e. that the relations~\req{kl} are
also the necessary conditions for their existence. 

The solution $\phi(x)$ with the period \req{per} has the total
wave number (that measured from the origin of Brillouin zone)
\eq{\tilde q\equiv Q-\frac{\phi_P}{P}\equiv 2\pi q=
2\pi\frac{k+l}{4k+3l} .}{qkl}
The values of $q$ allowed by conditions~\req{kl} form a 
Farey tree, shown in Fig.~\ref{farey} for the wave numbers 
between $q=1/3$ ($k=0$, $l=1$ and $P=3$) and $q=1/4$ ($k=1$, $l=0$
and $P=4$). Thus, already at this introductory stage of the analysis
we conclude that the model~\req{free} has the phase diagram with
branchings between neighboring commensurate configurations equivalent
to those of ANNNI models~\cite{selke1,selke2}. 

The  periodic solutions $q=1/3$ ($k=0$, $l=1$ and $P=3$) and 
$q=1/4$ ($k=1$, $l=0$ and $P=4$) in the Farey tree of Fig.~\ref{farey}
are the basic commensurate configurations, belonging to the Umklapp terms
of third and fourth order respectively.
The wave numbers at lower levels of the Farey tree from
Fig.~\ref{farey} represent higher order commensurate solutions
which correspond to all positive values of $k$ and $l$. 
They are situated between two main resonances in the Poincar\'{e} cross 
section shown in  Fig.~\ref{poinc}.
Note that for small values of $B$ and $C$ their positions in the phase 
space (Fig.~\ref{poinc}a) perfectly match positions in the Farey
tree (Fig.~\ref{farey}). As the parameters $B$ and $C$ further increase, 
the positions of periodic solutions that are embedded in chaotic layers 
become slightly intermixed since there are no more KAM tori between
two resonances that restrict their positions in phase space
(Fig.~\ref{poinc}c).  We do not 
include the parts of the Farey tree belonging to negative values of 
$k$ and/or $l$ since, as it will become clear later, they are not
thermodynamically stable for $0 < \delta_4  < \pi/6$ and $B, C > 0$.

In the next step we specify boundary conditions for a particular
periodic solution $\phi_{kl}(x)$. Since every periodic solution
possesses at least two inflection points, we chose one of them,
$x_0$, to be the initial point of integration, i.~e. the left end
point of one period. Thus $\phi_{kl}''(x=x_0)=0$. The boundary
conditions now read
\begin{eqnarray}
 \phi_{kl}(x=x_0 + P) &=& \phi_{kl}(x=x_0)+\phi_P , \nonumber \\
 \phi_{kl}'(x=x_0 + P) &=& \phi_{kl}'(x=x_0) . 
 \label{eq:percond}
\end{eqnarray}
Since the values of of $P$ and $\phi_P$ follow from the choice
of integers $(k,l)$ (Eqs.~\req{per}), it remains to establish
the connection between the other three parameters, $x_0$, 
$\phi_{kl}(x=x_0)$ and $\phi_{kl}'(x=x_0)$, which figure in the
conditions~\req{percond}. As it follows from the EL 
equation~\req{el} with $x=x_0$,
\eq{ 3B\sin{[3\phi(x_0) + 3(\frac{\pi}{6}-\delta_4)x_0]} +
     4C\sin{(4\phi(x_0) - 4\delta_4 x_0)}=0 ,}{x0el}
$x_0$ and $\phi(x_0)$ are not independent. Even more, the numerical 
experience suggests that for a given periodic solution both $x_0$
and $\phi(x_0)$ do not vary as we change $B$ and/or $C$, i.~e. that
Eq.~\req{x0el} in fact decomposes into two conditions, 
\begin{eqnarray}
 3\phi(x_0) &+& 3(\frac{\pi}{6}-\delta_4)x_0 = M\pi , \nonumber \\
 4\phi(x_0) &-& 4\delta_4 x_0= -N\pi ,
 \label{eq:x0phi0}
\end{eqnarray}
where $M$ and $N$ are integers. This means that $x_0$ and
$\phi(x_0)$ may have values 
\eq{ x_0=\frac{1}{2}(4M+3N) }{x0}
and
\eq{ \phi(x_0)=\delta_4 x_0 - \frac{\pi}{4}N .}{phix0}
These relations would allow for $2P$ values of $x_0$ and an
infinite number of values for $\phi(x_0)$ (for a general value of
$\delta_4$). The further analysis of symmetry properties of the
problem~\req{el}, as well as the numerical insight, however
indicate that for any choice of periods  $P$ and $\phi_P$ this
enumerable set is highly degenerate and reduces to only two
distinct (nondegenerate) solutions. The convenient choices of
$x_0$ and $\phi(x_0)$ characterizing these solutions for various
combinations of odd and/or even values of integers $k$ and $l$
are listed in Tab.~\ref{table1}. 

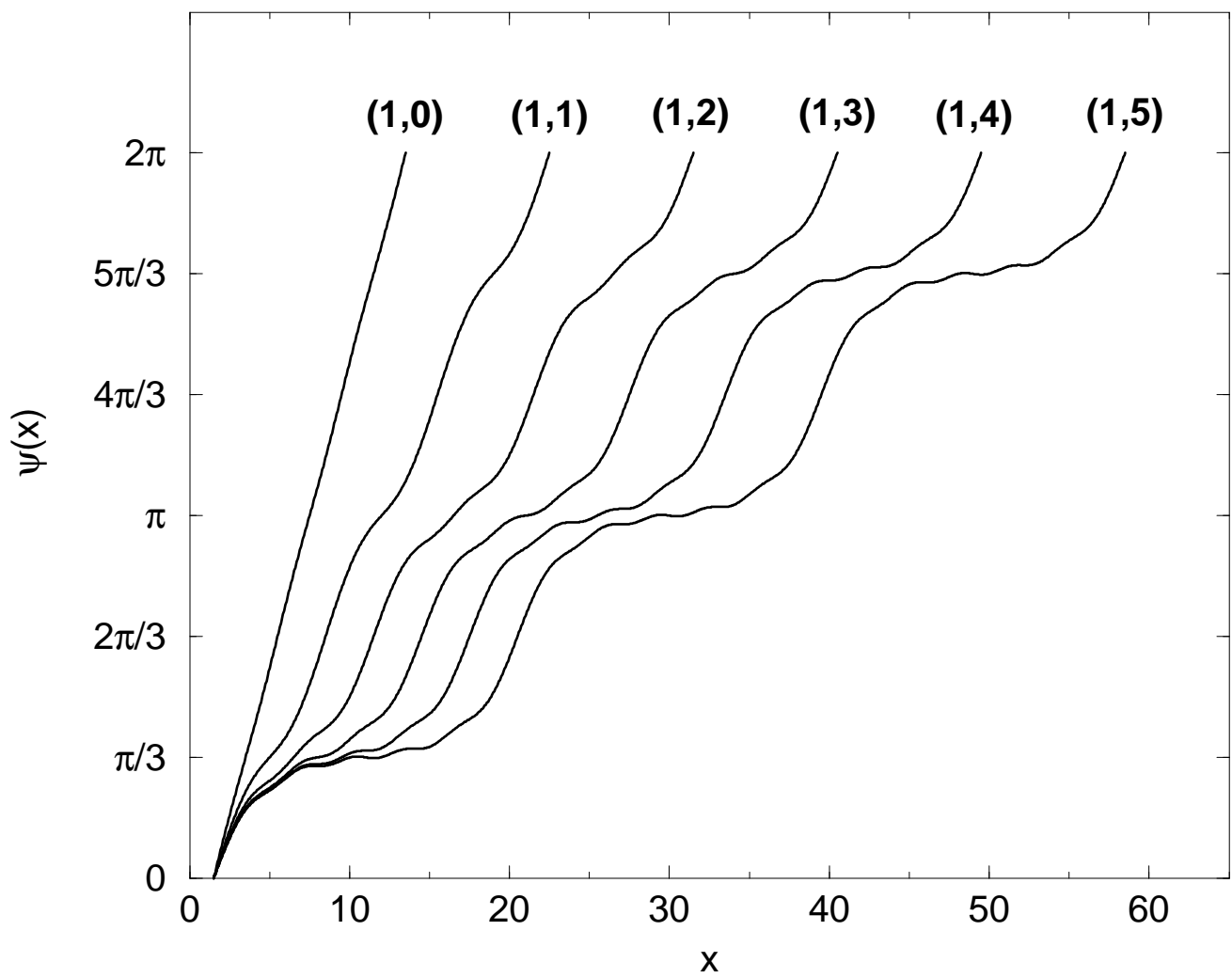
\begin{figure}
\caption{The periodic solutions $\psi(x)$ from the class $(1, l)$.
The parameters are: $B=0.02$, $C=0.02$, $x_0=1.5$, $\psi(x_0)=0$.}
\label{persol}
\end{figure}

The above analysis simplifies drastically the numerical procedure, 
since after specifying the parameters $k$, $l$, $x_0$ and $\phi(x_0)$, 
the determination of a given periodic solution follows from the 
variation of the single remaining parameter $\phi'(x_0)$. In
accomplishing this procedure it appears convenient to eliminate,
by transforming the variable $\phi(x)$, the explicit 
$x-$dependence from one of Umklapp terms in the EL equation
\req{el}, and to keep this dependence in the term with a weaker
amplitude. Thus for $B$ larger than $C$ we use the variable
$\psi(x)$ (Eq.~\req{psi}) instead of $\phi(x)$, so that the EL equation 
\eq{ \psi''+3B\sin{(3\psi)}+4C\sin{(4\psi-\frac{2\pi}{3}x)}=0 ,}{elpsi}
and its solutions $\psi(x)$ do not depend on $\delta_4$. The
corresponding free energy then acquires a $\delta_4$-dependent
term in the form of Lifshitz invariant,
\eq{ F=\int dx\{\frac{1}{2}[\psi'-
	(\frac{\pi}{6}-\delta_4)]^{2}+
	B\cos{(3\psi)}+C\cos{(4\psi-\frac{2\pi}{3}x)}\} ,}{freepsi}
which simplifies the calculation of the $\delta_4$-dependence of the
averaged free energy for any particular periodic solution of the EL 
equation. As is visible in Fig.~\ref{persol}, the form of periodic 
solutions resembles to that of multisoliton solutions of 
simple sine-Gordon model (still, note the slight modulation of
commensurate regions, i.~e. between discommensurations)
When $C$ is larger than $B$, it is appropriate to introduce 
an analogous variable which makes the $C$-term $x-$independent,
namely $\chi(x)=\phi(x)-\delta_4 x$. Again, the corresponding
EL equation does not depend on $\delta_4$. The boundary conditions
have to be modified correspondingly for both transformations.

Although the steps described above greatly simplify the numerical
procedure, the finite difference method poses the limitations on the
computer memory and time which do not allow us to calculate solutions
with periods well above 100. Note in this respect that the nonlinearity
of EL equation~\req{el} or~\req{elpsi}  forces us to use about 1000
mesh points per period in order to get solutions which are reliable
enough.

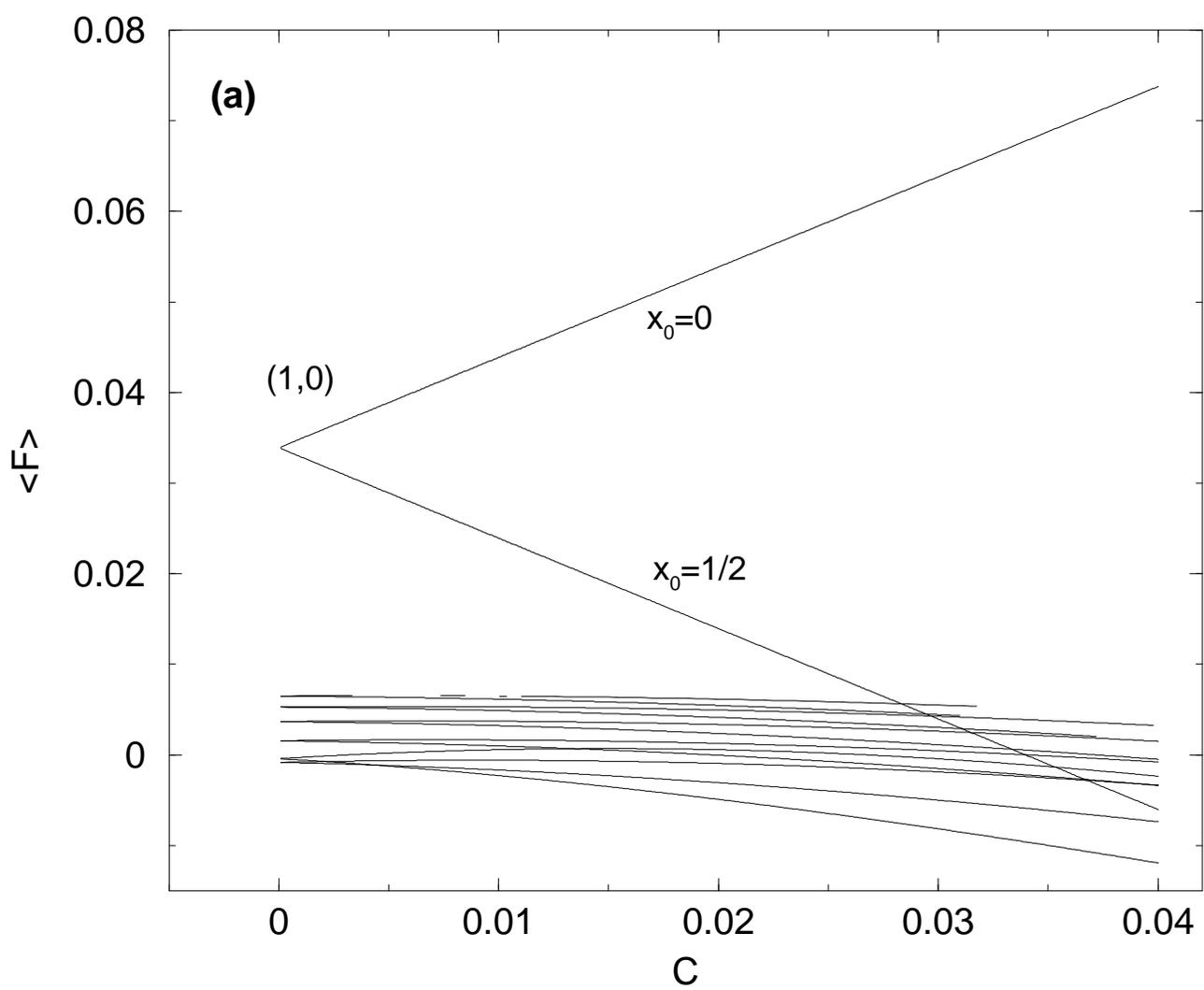
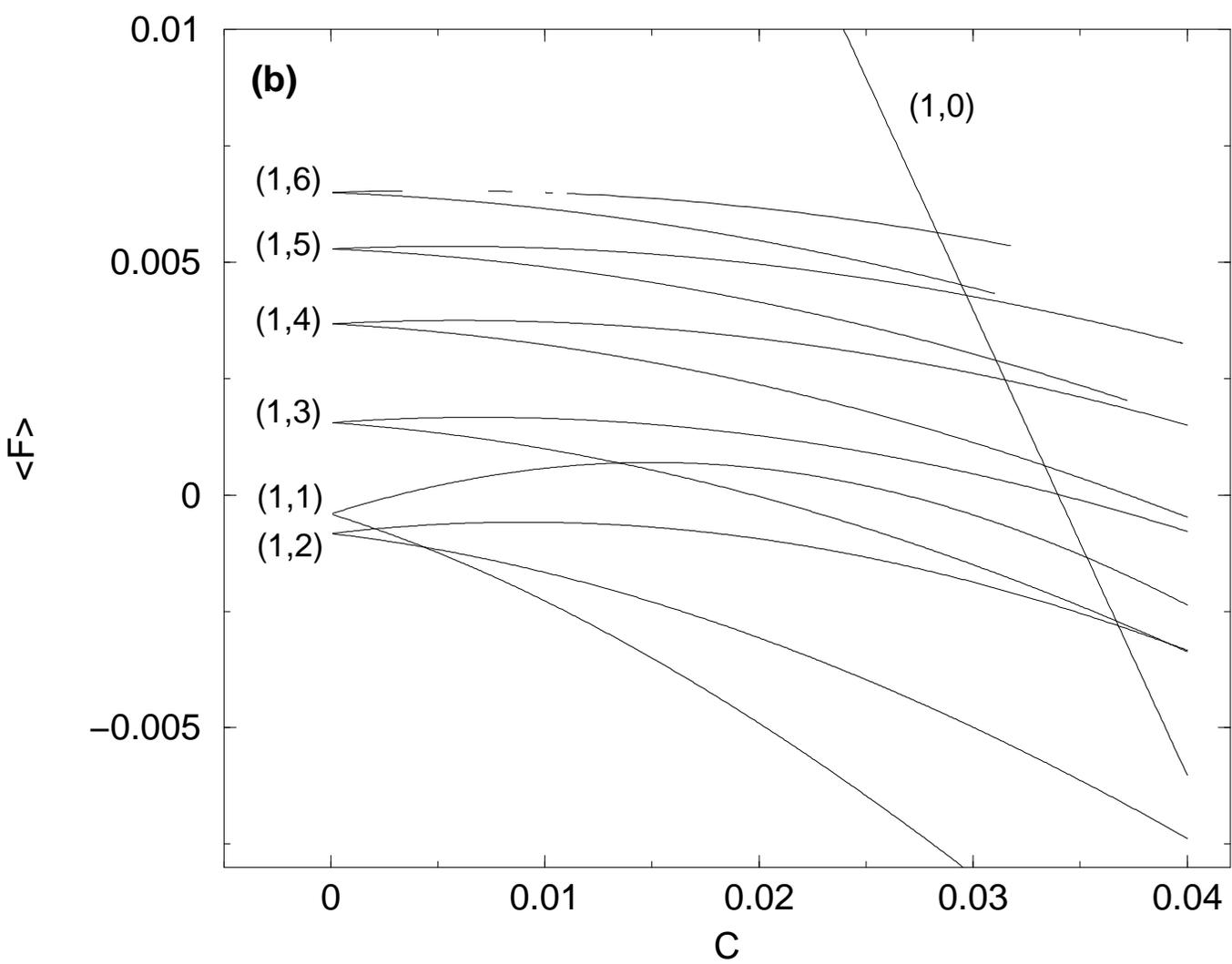
\begin{figure}
\caption{Average free energies of the periodic solutions
from the class $(1, l)$ as the function of $C$ for  
$B=0.02$ and $\delta_4=\pi/12$. 
Figure (b) is the enlarged detail of the figure (a) with energies
lower than $0.01$. Solutions with lower average
free energies are those from the second rows in Tab.~\protect{\ref{table1}}.
Note from the figure (b) that e. g. the upper solution $(1,6)$ 
does not exist for few subranges of the values of parameter $C$,
and that both solutions from this class cease to exist for $C > 0.03$.}
\label{split}
\end{figure}

Periodic solutions of EL equation~\req{el} show several interesting 
properties which are important for analysis of phase diagrams.
We notice that for some values (or ranges of values) of parameters
$B$ and $C$ one of the two periodic solutions with the same values
of $k$ and $l$ from Tab.~\ref{table1} ceases to exist 
(see Fig.~\ref{split}). In general the solutions with the lower value of
averaged free energy are more robust to this disappearance. We do
not go into a closer analysis of this effect, but only indicate that
it seems to be closely connected with the destruction of KAM tori as
$B$ and $C$ increase.

Another interesting property of periodic solutions is the splitting 
in averaged free energies of two solutions with the same values of
$(k,l)$ (see Fig.~\ref{split}). As the parameter $C$ gradually
increases from zero, while keeping $B$ fixed, values of the
difference between these two energies increases, thus making one
periodic solution more and more thermodynamically favorable with
respect to the other. This splitting is larger for the solutions
with smaller periods. The qualitative consequence is that such
solutions participate over greater and greater parts of the phase
diagram as parameters $B$ and $C$ increase.

For the calculation of quasiperiodic and chaotic trajectories we use
standard, Adams or Runge-Kutta-Merson, methods for an initial value 
problem. Quasiperiodic trajectories, as a building objects of
KAM tori, are orbitally stable~\cite{frad}.
Chaotic orbits, although certainly orbitally unstable, are diffusive 
through all the corresponding chaotic layer in the phase space, so that 
by picking one of them we get practically 
the averaged free energy for all chaotic solutions in that layer.
Thus, in order to calculate the averaged free
energies of quasiperiodic and chaotic solutions we chose initial values
by random (probability of picking periodic solution instead of 
quasiperiodic or chaotic ones is equal to zero), and carry out the
integration as long as the accuracy is satisfying. 
The fact that the averaged free energy of quasiperiodic and chaotic 
solutions can be determined only to a limited accuracy was already
pointed out by Fradkin et. al~\cite{frad} who estimated
the degree of accuracy for a given type of solution. 

The estimation of the common averaged free energy of chaotic solutions 
within a given layer can be done as follows~\cite{vlado}. The average 
value of Umklapp terms in the expression~\req{freepsi} is zero since
these terms contain trigonometric functions with an argument which
chaotically (randomly) varies with $x$. For the fourth order Umklapp
term $\cos{(4\psi-\frac{2\pi}{3}x)}$ we have
\eq{ \langle \cos{(\frac{2\pi}{3}x)}\cos{(4\psi)} \rangle = 
     \langle \sin{(\frac{2\pi}{3}x)}\sin{(4\psi)} \rangle = 0 ,}{ave-u}
while for the third order Umklapp term we have an average of
$\cos{3\psi}$ which is also zero. The averaged free energy is thus
given by the integral of the gradient term
$\frac{1}{2}[\psi'-(\pi/6-\delta_4)]^2$. The latter depends on the
position and the width of chaotic layer in the phase space, i.~e. only
on the dependence of $\psi'$ on $x$ along the trajectory in this layer.

In order to determine a solution with the lowest average free energy
we follow solutions (periodic, quasiperiodic and chaotic) with initial
conditions that belong to the line $(\psi(x_0)=0,\psi'(x_0))$ in the phase
space $(\psi,\psi')$ (Fig.~\ref{poinc}), and compute their average
free energies. For small values of $B$ and $C$, periodic and
quasiperiodic solutions are regularly arranged in the phase space
(Fig.~\ref{poinc}a), with hardly distinguishable average free energies
(Fig.~\ref{fpsid}a). In order to show that the solution with the lowest
free energy is periodic, we follow downwards the branch of the Farey
tree (Fig.~\ref{farey}) which starts at the point with the averaged
free energy lower that those for neighboring points above and below
this point. It is a numerical evidence that the average free energies
increase (and tend to some finite value) as we go down through
successive branch points, i.~e. through the solution with larger and
larger periods. Quasiperiodic solutions can be regarded as asymptotic
limits of series of periodic solutions defined by successive branchings
in the Farey tree, in which the period and the phase increment tend to
infinity (but with a finite, irrational, value of $q$). The averaged
free energies of quasiperiodic solutions thus should be equal to the
limiting values of averaged free energies at a given branch, which are,
as is argued above, higher than the averaged free energy of the starting
periodic solution. Since this argument is based on numerical
calculations, it cannot be extended to very small values of $B$ and $C$
for which the solution with the lowest free energy, as well as the
solutions at the accompanying branch in the Farey tree, have too large
periods.

In the range of intermediate values of $B$ and $C$ (Fig.~\ref{fpsid}b)
there are intervals of initial conditions in which quasiperiodic
solutions disappear, and only chaotic and periodic solutions are
present. The chaotic layers can be easily recognized in the Poincar\'{e}
cross section (Fig.~\ref{poinc}b). The average free energies of periodic
solutions then look as needle-like minima immersed in the average free
energy of chaotic layers, represented by plateaux in Fig.~\ref{fpsid}b.
Finally, for large values of $B$ and $C$ (Fig.~\ref{fpsid}c) for which
the Chirikov criterion~\req{chirc} is fulfilled, there remains a single
chaotic layer between two resonance domains (Fig.~\ref{poinc}c),
while the number of existing periodic solutions gradually decreases
as $B$ and $C$ increase. Since there remains a finite number of 
corresponding well defined needle-like minima, it is sufficient to
limit the numerical calculations to the search for existing periodic
solutions, and to find out among them the solution that has the lowest
average free energy.


\section{Phase diagram}

We have argued in the previous section that the configurations
with minimal average free energy are among periodic
solutions of EL equation~\req{el}. Before presenting
results of numerical calculations which confirm this expectation, 
we briefly discuss the parameters present in the model~\req{free}.

The parameters $B$ and $C$ depend on external conditions, most
usually on temperature and pressure. As it was mentioned in Sec.~II, 
they depend on the amplitude of the order parameter linearly and 
quadratically respectively. At temperatures closely below $T_I$,
the temperature of phase transition from the disordered to the
incommensurate phase, the ratio $B/C$ is proportional to 
$(T_I-T)^{-1/2}$. A more complete insight into the temperature
dependence of the order parameter, and of the ratio $B/C$ as well,
in the wider temperature range below $T_I$, can be obtained from the
neutron scattering, NMR and similar experimental data for particular
materials (e.~g. references~\cite{blin1,blin2}).
As for the pressure dependence of $B$ and $C$, it can be specified
only after the insight into the microscopic model for a particular
material on which the Landau theory is based. The parameter $\delta_4$
also might be temperature and/or pressure dependent. Usually, in a
concrete physical situation certain dependences may be regarded as
dominant. E.~g., when temperature varies and pressure is constant
$\delta_4$ can be often regarded as constant, while $B$ and $C$ are
temperature dependent. Having this in mind we simplify the further
discussion by keeping one of the parameters fixed and concentrating
on phase diagrams in remaining two-dimensional parameter subspaces. 

The role of the parameter $\delta_4$, the position of the instability
with respect to the wave number of the fourth order commensurability,
is expressed through the Lifshitz invariant $\delta_4 \psi'(x)$ in
Eq.~\req{freepsi} which favors the incommensurate ordering. On the
other side two Umklapp terms favor commensurate orderings with their
respective wave numbers. For $\delta_4\rightarrow 0$, and fixed values
of parameters $B$ and $C$, the Umklapp term of the fourth order dominates 
with respect to that of the third order, and the thermodynamically stable 
periodic solution is expected to have the wave number $q_0=1/4$. On
the same footing, for $\delta_4$ near $\pi/6$ (i.~e. for 
$\delta_3 \rightarrow 0$) the stabilization of the modulation with
$q_0=1/3$ is preferred. For $0 < \delta_4 < \pi/6$ we expect that some
other higher-order wave numbers of modulation become thermodynamically
stable and that they follow the order specified by the Farey tree from 
Fig.~\ref{farey}.

Let us now fix the parameter $B$ and allow for the variation of the
parameters $\delta_4$ and $C$. For a particular value of $C$ we find
periodic solutions of the EL equation~\req{elpsi} by following the
steps from Sec.~III, and calculate their average free
energy~\req{freepsi} for a relevant range of values of the parameter
$\delta_4$. Then we determine a solution which is the absolute
minimum of the average free energy for a given value of $\delta_4$,
and in particular the isolated values of $\delta_4$ at which first
order phase transitions take place since two (or more) configurations 
are simultaneously absolute minima of the free energy. Varying also
systematically the parameter $C$ we obtain the phase diagram, as
shown in Fig.~\ref{cd4b02} for $B=0.02$. All lines in this diagram 
represent the phase transitions of the first order between the periodic 
configurations with different wave numbers (which are denoted only for 
few dominant phases in the diagram). Note that the Chirikov
line~\req{chirc} is at $C\approx 0.0145$, and that below
$C\approx 0.01$ there is a proliferation of configurations with
commensurabilies of higher and higher orders. The absence of these
configurations at larger values of $C$ is mostly due to the fact that, 
although they exist as solutions of the EL equation, their average free 
energies are too high in comparison to those of the solutions with lower 
commensurabilities. In addition, some periodic solutions simply cease to
exist as $C$ (or $B$) increase, as shown in Fig.~\ref{split}.
Note also that only one of two different classes of periodic solutions 
with the same values of $k$ and $l$ participates in the phase diagram in 
Fig.~\ref{cd4b02}, i.~e. that characterized by the initial conditions 
from the second rows (depending on evenness and oddness of integers $k$ 
and $l$) in Tab.~\ref{table1}. Still, we find out numerically that
the average free energies for two different solutions with the same
$(k,l)$ may change order, i.~e. that the solutions from the first rows
in Tab.~\ref{table1} may have lower average free energy than those
from the second rows provided they are of rather high commensurability.
Thus, it is somewhat surprising that in the phase diagram in
Fig.~\ref{cd4b02} the periodic solutions with only one type of
boundary conditions prevail. We shall come back to this point later
in Sec.~V.

In addition to the phase diagram, we plot in Fig.~\ref{devb02}
the corresponding staircase, i.~e. the wave number of the stable
configuration {\em vs} parameters $C$ and $\delta_4$. As long as
$C$ is not very small there is a finite number of steps, i.~e. we
obtain the so-called harmless staircase, introduced by Villain
and Gordon~\cite{vill}. We stress that the most
interesting property of the phase diagram from Figs.~\ref{cd4b02}
and \ref{devb02}, the presence of a finite (small) number of stable 
commensurate configurations, is encountered in the regime of rather
high values of parameters $B$ and $C$. The phase
portrait of the EL equation~\req{el} is then almost everywhere
chaotic (Fig.~\ref{poinc}c) and there are no more quasiperiodic
solutions between two resonances. By increasing further the 
values of parameters $B$ and $C$ one eventually comes to the phase 
diagram in which only two main commensurate phases ($q=1/3$ and
$q=1/4$) take place. 

Another possible presentation of the phase diagram is that with a fixed
value of the parameter $\delta_4$ and with varying parameters $B$
and $C$. It is presumably closer to usual physical situations in
which only weak temperature and pressure dependences of $\delta_4$ are
expected. The construction of the $(B,C)$ phase diagram is however
computationally more demanding, since one has to look for the
solution with the lowest average free energy within a set of
solutions for given values of $B$ and $C$, i.~e. one has to 
calculate the whole set of periodic solutions of the EL equation
[Eq.~\req{el} or \req{elpsi}] for each point in the two-dimensional
phase diagram. To this end we use a mesh of points which is dense enough
in the $(B,C)$ plane, and determine the solution with the lowest
average free energy at each point. The phase diagram obtained in
this way for $\delta_4=\pi/12$ is shown in Fig.~\ref{d4fix}. Note
that again only configurations with rather low orders of
commensurability, i.~e. with small values of parameters $(k,l)$,
are present above the Chirikov line (Eq.~\req{chirc}),
while below this line the diagram is more complex since
a great number of first order transitions takes place within a
small part of the phase diagram. 


\section{Conclusion}

The most important conclusions  of the above analysis follow from
the thermodynamic phase diagram obtained in the regime of comparable
strengths of two Umklapp terms included into the Landau
expansion~\req{free}. At first we emphasize that only one type of
solutions of the corresponding EL equation, namely periodic
configurations, participates in the phase diagram. Furthermore,
all phase transitions between successive commensurate phases are
of the first order, so that the wave number of ordering follows
a harmless staircase with a finite number of steps. The examples
of such phase diagram, namely series of successive lock-in
commensurate-commensurate transitions with accompanying effects which
characterize first order transitions~\cite{cumm}, are often encountered
in particular materials. Here we focus our attention on few well known
examples. 

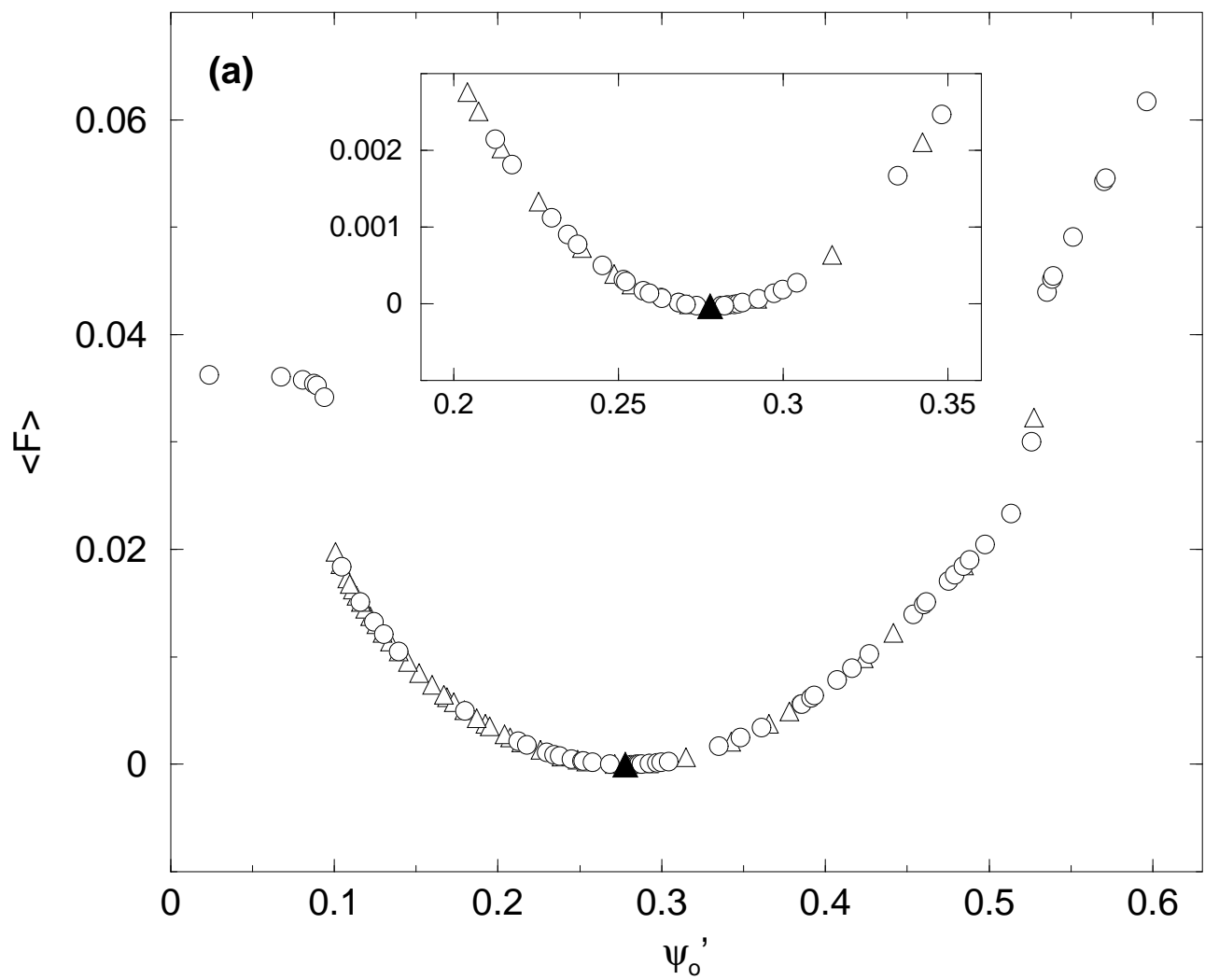
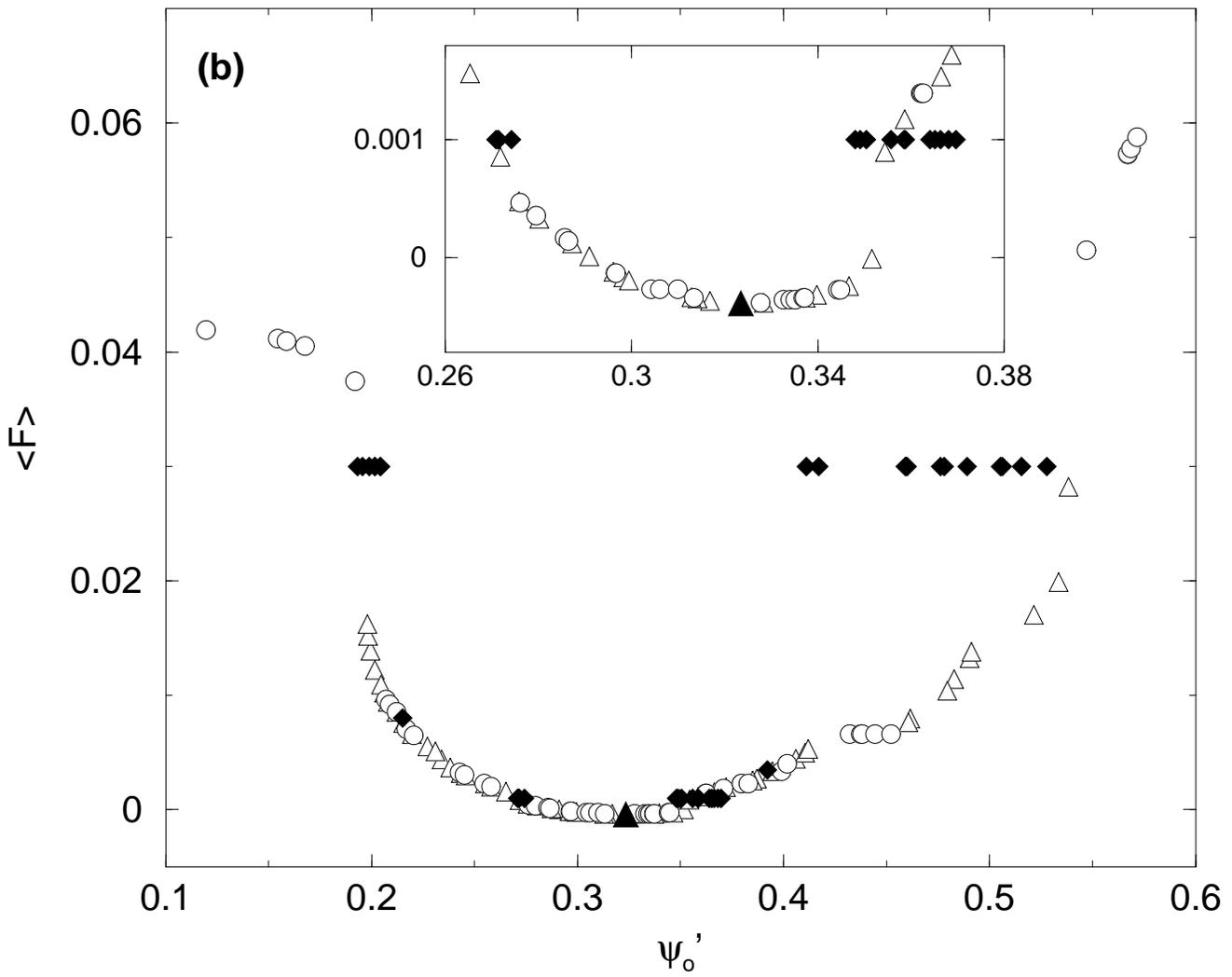
\begin{figure}
\caption{Average free energy {\em vs} $\psi_0'$ of periodic 
($\triangle$), quasiperiodic ($\bigcirc$) and chaotic (solid 
$\Diamond$) solutions, for $B=0.002$, $C=0.002$ (a), $B=0.008$, 
$C=0.006$ (b), $B=0.02$, $C=0.02$ (c), and $\delta_4=\pi/12$,
$x_0=1.5$. The $(k,l)$ indices for the periodic solution with 
the lowest average free energy (solid $\triangle$)
are $(3,4)$ in the figures (a) and (b), and $(1,1)$ in the figure (c).
The insets in figures (a) and (b) are enlarged neighborhoods of the
free energy minima.}
\label{fpsid}
\end{figure}

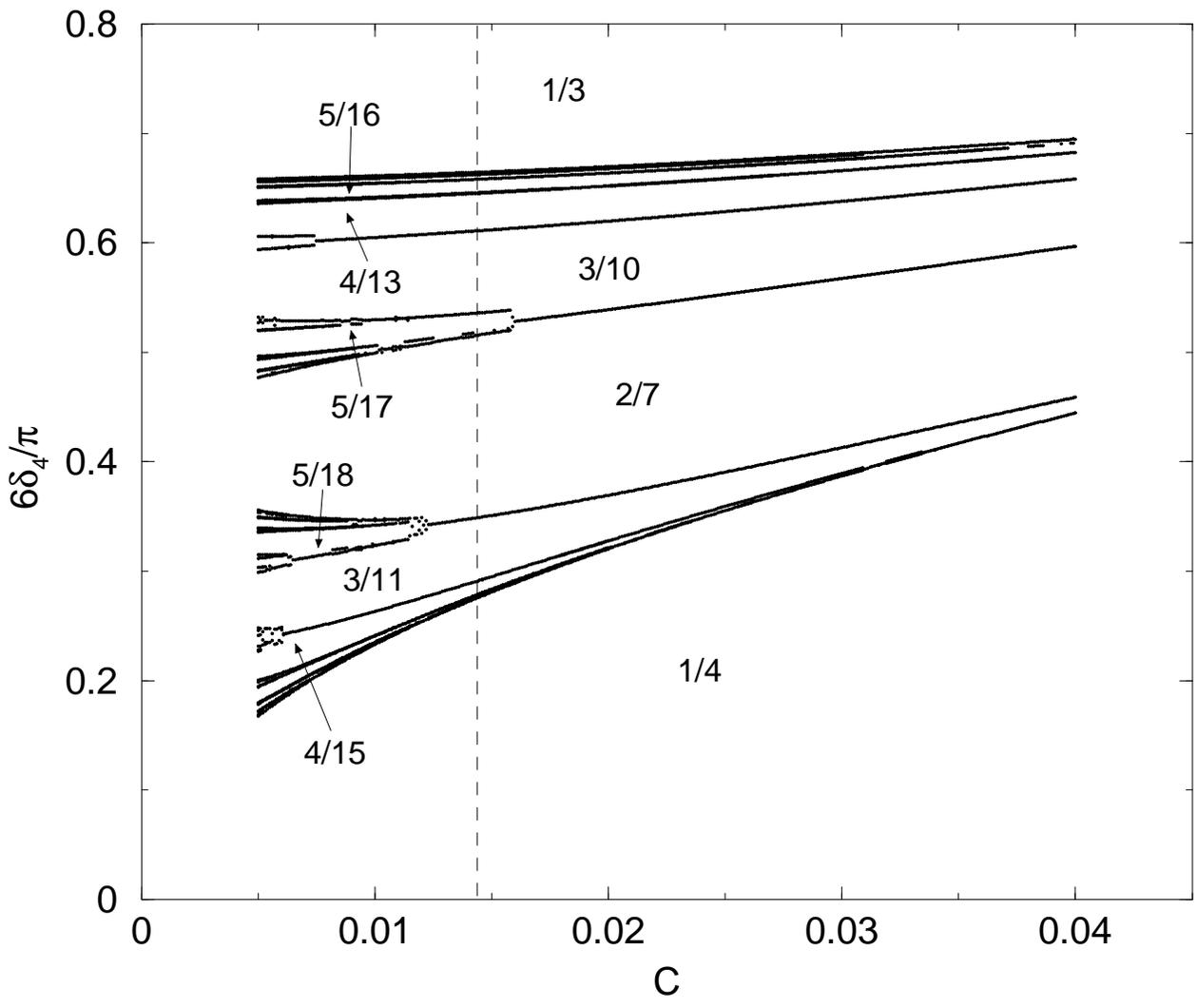
\begin{figure}
\caption{Phase diagram in the $(C,\delta_4)$ plane for $B=0.02$. 
The numbers in the figure are periods of some stable commensurate 
phases. The dashed line at $C\approx 0.0145$ represents the 
Chirikov criterion~\protect{\req{chirc}}.}
\label{cd4b02}
\end{figure}

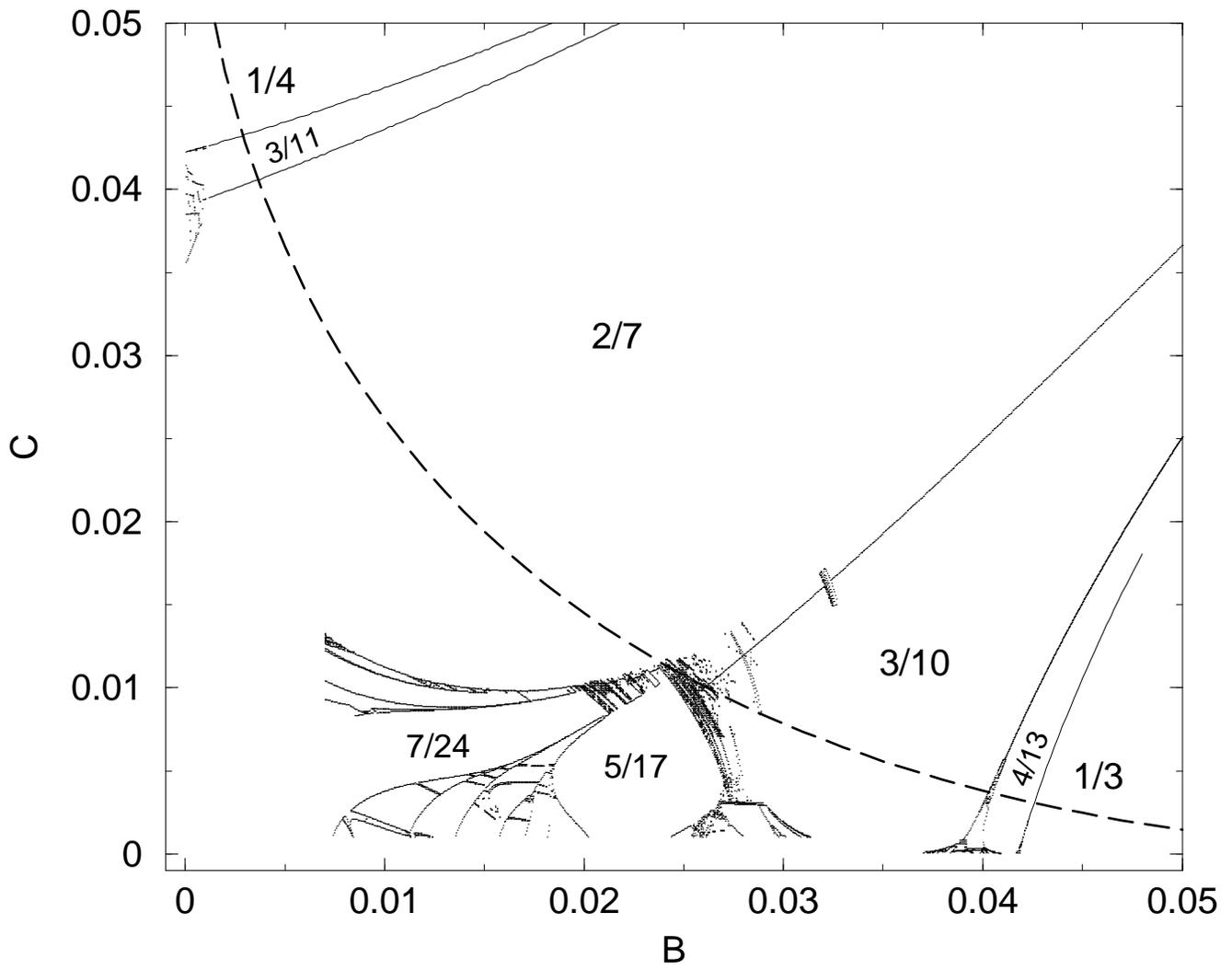
\begin{figure}
\caption{Phase diagram in the $(B,C)$ plane for $\delta_4=\pi/12$.
The numbers in the figure are periods of some stable commensurate 
phases. The dashed curve represents the Chirikov
criterion~\protect{\req{chirc}}.} 
\label{d4fix}
\end{figure}

One of the most studied type of materials are A$_2$BX$_4$ compounds,
among which we take Rb$_2$ZnBr$_4$ as a prominent representative.
Early neutron diffraction measurements~\cite{iizu,pate,iizu2} of
the temperature variation of modulation wave number revealed the
existence of several higher-order commensurate phases. The more
complete pressure-temperature phase diagram followed from various
subsequent data, in particular again from the neutron diffraction
measurements of Parlinski et al.~\cite{parl1,parl2}. It resembles
to a great extent to our phase diagrams from Figs.~\ref{cd4b02}
and \ref{d4fix}. Note also that the phenomenological formula for
wave numbers of observed commensurate phases introduced 
by Parlinski et al.~\cite{parl1} coincides to our expression for Farey 
tree~\req{qkl}, which is, as is shown in Sec.~III, inherent to the
model~\req{free}. Harmless staircase are clearly seen in e.~g. pressure
variation of the wave vector for a fixed temperature~\cite{parl1}, with
steps going as $1/3$, $7/24$, $2/7$ and $1/4$ by increasing pressure.
They are accompanied by hysteresis in pressure and temperature runs,
which are particularly strong when only a few steps appear in the phase
diagram. This corresponds to the regime of rather high values of
parameters $B$ and $C$, in which the phase diagram contains only few
commensurate phases and the average free energy of chaotic plateau is
well above average free energies of periodic solutions
(Fig.~\ref{fpsid}c).

Existing theoretical approaches to the (in)commensurate orderings in
A$_2$BX$_4$ compounds, in particular to the appearance of series of
commensurate phases, are mostly phenomenological, based either on
Landau expansions~\cite{ishi} or on the discrete models of competing
local interactions~\cite{yama,chen,janssen2}. The justification for the
continuous Landau models, which are generally appropriate to weak 
coupling systems, comes from many experimental indications, starting
from the early neutron scattering data~\cite{iizu,pate,iizu2}, showing
a well-defined dispersion curves for collective modes with distinct
soft-mode minima. However, the previous analyses of Landau models
were restricted to purely sinusoidal modulation, and, as such were not 
able to explain the appearance of phases with commensurabilities 
of orders higher than three or four. It was therefore 
proposed that such phases appear due to the presence of Umklapp terms 
of higher orders in the free energy expansion~\cite{mashi,marion,parl3}.
This explanation, which is based on the assumption that distinct 
commensurate stars of wave vectors are necessary for the stabilization
of, presumably sinusoidal, phases with corresponding wave 
vectors~\cite{mashi,marion,parl3}, is not convincing since the
Umklapp terms of order higher than four are expected 
to be negligible in weakly coupled systems with a displacive order. 
 
For these reasons the more recent attempts turned again towards another 
type of approaches, those which assume strong couplings, so that 
the lattice discreteness has to be taken into account. Originally 
the sequences of IC-C and C-C phase transitions were within this scheme
interpreted in terms of the FK model as devil's staircase dependences
of the wave number of ordering, i.~e. as dense sequences of second order 
phase transitions~\cite{aubr}. However, observed staircase
rarely resemble, even within experimental limitations, to the dense 
devil's staircase. Beside, phase transitions between successive C
phases are usually of the first order.
The phase diagrams which are closer to experimental findings
may be however obtained by various extensions of the basic FK model,
e.~g. by including an additional harmonic potential~\cite{grif,yokoi}.
Also, the more complex models of competing interactions, e.~g.
DIFFOUR~\cite{janssen1} and ANNNI~\cite{fisher,selke1,yama,selke2}
models, as well as models that assume two critical modes per lattice
site~\cite{chen,janssen2}, are particularly successful in describing
the phase transitions in A$_2$BX$_4$ compounds. Within some of these
models (e.~g. Refs.~\cite{chen} and~\cite{janssen2}) one also obtains
the first order phase transitions between configurations having the
same wave numbers but different symmetries. As was already stated in
Sec.~IV, this is not the case within the model~\req{free}, i.~e.
although the EL equation~\req{el} may possess two types of solutions
with the same periods, only one type of solutions participates in the
phase diagram.

The present analysis again starts from the minimal Landau expansion 
(with terms up to the fourth order), but takes into consideration all 
solutions of the corresponding EL equation.
In particular it indicates that the theoretical approach~\cite{iizu},
proposed together with the first neutron scattering measurements on 
Rb$_2$ZnBr$_4$, might be essentially sufficient for the understanding of 
complex phase diagrams in A$_2$BX$_4$ materials. 
The more detailed analysis which takes into account some additional
aspects, like the couplings to the homogeneous polarization and strain 
which appear in some materials as secondary order parameters, will be 
done elsewhere.

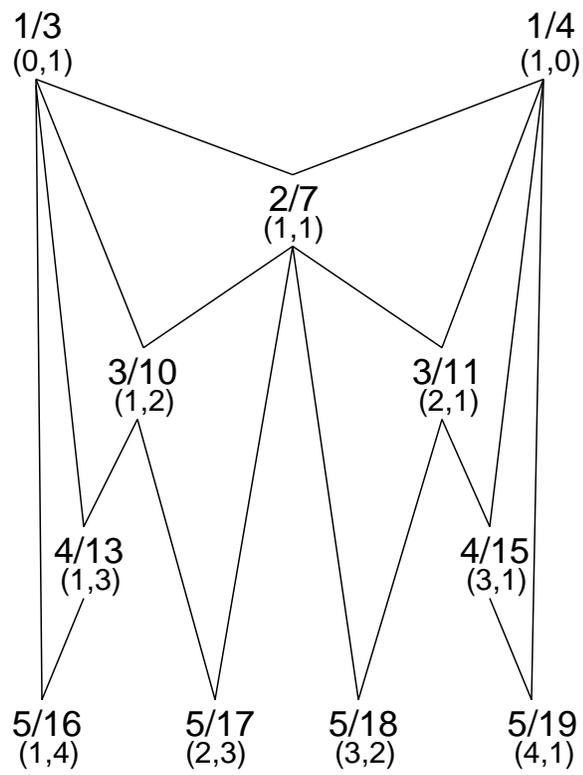
\begin{figure}
\caption{Farey tree for wave numbers $q$ defined by
Eq.~\protect{\req{qkl}}.}
\label{farey}
\end{figure}

Betaine-calciumchloride-dihydrate (BCCD), together with its deuterated
version D-BCCD, belongs to the second type of intensely studied materials
with the commensurate lock-ins. It shows an exceptionally rich 
staircase going from $q=1/3$ down to $q=0$, with numerous intermediate 
steps of higher orders~\cite{kiat,chav1,chav2}.
A closer insight into the region of phase diagram with the wave number
between $q=1/3$ and $q=1/4$ shows that only upper right triangle of the
Farey tree from Fig.~\ref{farey} is realized, i.~e. the phase diagram is
mostly covered by wave vectors close to $q=1/4$, and not by those close
to $q=1/3$. This sequence of IC-C transitions was successfully interpreted
within various discrete models with competing interactions, e.~g. in 
Refs.~\cite{kappler,neubert94}.
Within the model~\req{free} such phase diagram corresponds to the
regime in which the fourth order Umklapp term dominates with respect to
that of the third order. Also, two types of extensions of our model 
may lead to the stabilization of commensurate phases with $q<1/4$.
Namely, one may allow for negative values of the parameter $\delta_4$,
or start with other Umklapp terms, e.~g. with those of the fourth
and fifth order, and pursue the analysis analogous to that of   
Secs.~III and IV.

We also mention some other materials that exhibit a sequence of IC-C
and C-C phase transitions between $q=1/3$ and $q=1/4$, but are not so
extensively studied as previous two examples. E.~g., D\'{e}noyer et
al.~\cite{deno} investigated NH$_4$HSeO$_4$ and its deuterated version
ND$_4$DSeO$_4$ by neutron diffraction, and found the harmless staircase
and first order phase transitions, accompanied by the coexistence of
several phases in the relatively wide range of temperatures.
A series of IC-C and C-C phase transitions are also observed in
BaZnGeO$_4$ in X-ray diffraction measurements by Sakashita et
al.~\cite{saka}, and in electron diffraction measurements by
Yamamoto et al.~\cite{yamamoto} which also provide dark field images
of discommensurations appearing in the vicinity of $q=1/3$ phase.
An example of particularly sharp transition from $q=1/3$ to $q=1/4$,
with a very wide temperature range of the coexistence of these two
commensurate phases, was found by Broda~\cite{broda} in
(NH$_4$)$_2$CoCl$_4$, the material that also belongs to A$_2$BX$_4$
family. 

The free energy~\req{free} is similar to that of Fradkin et
al.~\cite{frad} who also studied continuum systems with competing
periodicities. The only difference between two expressions is the
absence of the factors 3 and 4 in front of the variable $\phi(x)$
in the cosine terms of the model~\cite{frad}. However, in contrast
to ours, the analysis carried out in Ref.~\cite{frad} is limited to
the close vicinity of the separatrices (and hyperbolic points) in
the phase space~\cite{chir}, i.~e. to the dilute soliton lattices.
Then the continuum model can be converted into a discrete mapping of
the FK type, analyzed in detail previously by Aubry~\cite{aubr}.
Our analysis covers the whole phase space, i.~e. all solutions of the
EL equation~\req{el}, and in particular the whole class of periodic
configurations. In particular, our thermodynamic phase diagram
(Figs.~\ref{cd4b02} and \ref{d4fix}) includes, in contrast to
that of Ref.~\cite{frad}, the most interesting part of the phase space,
namely that between two resonances (i.~e. sets of hyperbolic points).

The model~\cite{frad} was the starting point in the 
investigation~\cite{erra} of the memory effects in systems with
IC modulations, based on the earlier proposition~\cite{lederer} 
that mobile defects might be responsible, by forming defect density 
waves, for the sensitivity of the IC ordering on the thermal history 
of crystal, observed e.~g. in thiourea~\cite{jamet2}.
Errandonea~\cite{erra} argued that the double sine-Gordon model,
with two lock-in potentials originating from the lattice the defect
density wave, is an appropriate description of this phenomena.

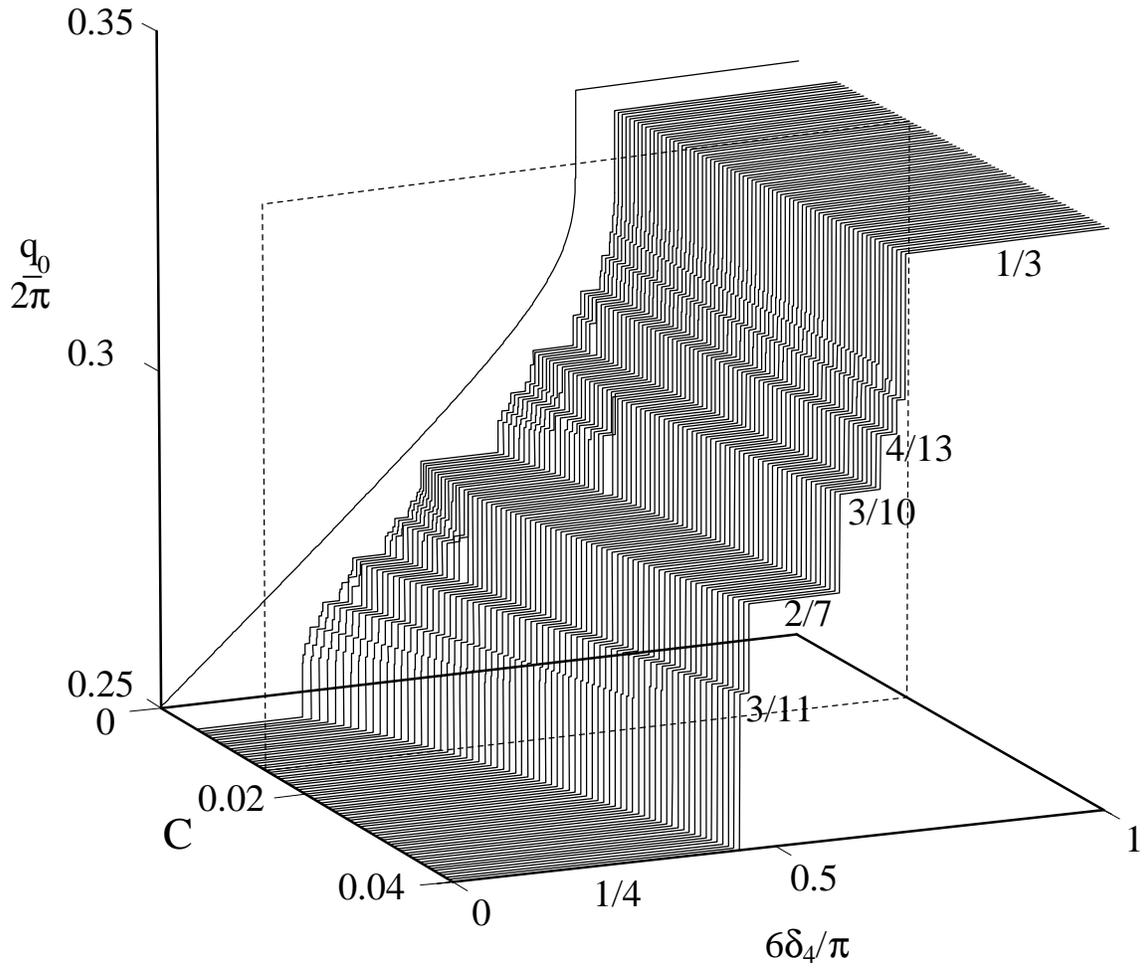
\begin{figure}
\caption{The wave number of modulation $q_0$ {\em vs} $C$ and $\delta_4$
for $B=0.02$. The dotted cross-section represents the Chirikov
criterion~\protect{\req{chirc}}.}
\label{devb02}
\end{figure}

The model~\req{free} provides the explanation of memory effects 
(together with thermal hystereses), without referring, in contrast to 
models~\cite{erra,lederer}, to defects as an intrinsic ingredient of 
the theory. At first, we note that the crossings of lines of first order 
phase transitions in Figs.~\ref{cd4b02}, \ref{devb02} and \ref{d4fix}
are accompanied by hystereses. Our preliminary analysis~\cite{dlb} 
indicates that these hystereses may be rather wide on e.~g. 
temperature scale. Furthermore, the present analysis of the
EL equation~\req{el} shows that periodic solutions, which constitute the 
phase diagram, are immersed as isolated points into the environment of 
chaotic configurations. This  environment prevents both, the continuous 
variation of the wave number of ordering and the
continuous phase transitions of the second and higher orders. The
average free energy of chaotic solutions from Fig.~\ref{fpsid}c
is the measure of the energetic barrier which the system has to overcome 
in order to pass from some periodic (metastable) configuration
to another one with lower free energy. This is expected
to be a common property of models which are nonintegrable (beside
being nonlinear), and have thermodynamically stable periodic
configurations isolated in the chaotic phase space~\cite{bari}. 

The memory effects are also observed in class II of IC
systems~\cite{cumm}. The detailed analysis of phase diagram for
this class~\cite{dana} led to the conclusion that the corresponding
phenomena seen in particular materials may be as well interpreted 
in terms similar to those presented above. However, it was also 
stressed ~\cite{dana} that defects
may have a secondary role as triggers which favor the stabilization 
of some domain patterns. This interpretation invokes neither the 
mobility of defects nor the formation of defect density waves.
The analogous secondary influence of defects on memory phenomena is 
expected also in presently investigated systems of class I.

Finally, let us mention a common problem that arises in the
analysis of continuous nonintegrable Landau models for uniaxial
systems of classes I~\cite{bjel} and II~\cite{dana}, in which 
periodic solutions have an essential role in the extremalization of
corresponding thermodynamic functionals. We remind that there is no
firm universal principle which would favor the thermodynamic stability
of the (meta)stable periodic configurations on account of other,
quasiperiodic or chaotic, solutions of EL equations. Some hints in this
direction for "autonomous" functionals (those for which the free energy
density does not depend explicitly on $x$) follow from the recently
derived general criteria~\cite{dana2} based on the additional
extremalizations (like e.~g. those involving boundary conditions).
However, these criteria cannot be directly applied to the present model
since the explicit $x$-dependence in Eq.~\req{free} introduces
fundamental singularities in the additional extremalizatons~\cite{dana2}.
Thus, the most important property of the phase diagrams from
Figs.~\ref{cd4b02} and \ref{d4fix}, their complete coverage by a
finite number of periodic configurations, still awaits for a deeper
understanding.

We acknowledge numerous useful discussions with V. Danani\'{c}. 
The work is supported by the Ministry of Science and Technology of 
the Republic of Croatia through the project no. 119201.



\begin{table}
\caption{The set of possible values of $x_0$ and $\phi(x_0)$ needed
for specifying boundary conditions of EL equation~\protect{\req{el}}.}
\begin{tabular}{ccccc}
$k$ & $l$ & $P$ & $x_0$ & $\phi(x_0)$ \\ \hline
odd & odd & odd &   0   &     0    \\ \cline{4-5}
    &     &     &   0   &   $\pi$  \\ \hline
even& odd & odd &   0   &     0    \\ \cline{4-5}
    &     &     &   0   &   $\pi$  \\ \hline
odd &even &even &   0   &     0    \\ \cline{4-5}
    &     &     & $1/2$ & $\frac{1}{2}\delta_4+\frac{\pi}{4}$ \\
\end{tabular}
\label{table1}
\end{table}


\end{document}